\documentclass[conference]{IEEEtran}
\IEEEoverridecommandlockouts
\usepackage{cite}
\usepackage{amsmath,amssymb,amsfonts}
\usepackage{algorithmic}
\usepackage{graphicx}
\usepackage{textcomp}
\usepackage{multirow}
\usepackage{enumitem}
\usepackage{xcolor}
\usepackage{caption,subcaption}
\usepackage{booktabs}
\usepackage{pbox}
\usepackage{tikz}
\usepackage{pifont}

\newcommand{\cmark}{\ding{51}}%
\newcommand{\xmark}{\ding{55}}%

\def\BibTeX{{\rm B\kern-.05em{\sc i\kern-.025em b}\kern-.08em
    T\kern-.1667em\lower.7ex\hbox{E}\kern-.125emX}}

\usepackage{fancyhdr}
\begin{document}

\title{PEFT-SER: On the Use of Parameter Efficient Transfer Learning Approaches For Speech Emotion Recognition Using Pre-trained Speech Models \\

\thanks{Thanks to the USC-Amazon Center for supporting this work.}
}

\author{\IEEEauthorblockN{Tiantian Feng}
\IEEEauthorblockA{\textit{Department of Computer Science} \\
\textit{University of Southern California}\\
Los Angeles, USA \\
tiantiaf@usc.edu}
\and
\IEEEauthorblockN{Shrikanth Narayanan}
\IEEEauthorblockA{\textit{Department of Electrical and Computer Engineering} \\
\textit{University of Southern California}\\
Los Angeles, USA \\
shri@ee.usc.edu}
}

\maketitle
\thispagestyle{fancy}

\begin{abstract}
Many recent studies have focused on fine-tuning pre-trained models for speech emotion recognition (SER), resulting in promising performance compared to traditional methods that rely largely on low-level, knowledge-inspired acoustic features. These pre-trained speech models learn general-purpose speech representations using self-supervised or weakly-supervised learning objectives from large-scale datasets. Despite the significant advances made in SER through the use of pre-trained architecture, fine-tuning these large pre-trained models for different datasets requires saving copies of entire weight parameters, rendering them impractical to deploy in real-world settings. As an alternative, this work explores parameter-efficient fine-tuning (PEFT) approaches for adapting pre-trained speech models for emotion recognition. Specifically, we evaluate the efficacy of adapter tuning, embedding prompt tuning, and LoRa (Low-rank approximation) on four popular SER testbeds. Our results reveal that LoRa achieves the best fine-tuning performance in emotion recognition while enhancing fairness and requiring only a minimal extra amount of weight parameters. Furthermore, our findings offer novel insights into future research directions in SER, distinct from existing approaches focusing on directly fine-tuning the model architecture. Our code is publicly available under: https://github.com/usc-sail/peft-ser.

\end{abstract}


\section{Introduction}
Speech emotion recognition (SER) \cite{busso2004analysis} aims to recognize the expressed emotional state of a speaker perceived based on their spoken utterances. A SER framework typically commences by converting speech signals to low-level descriptive (LLD) features, such as prosodic features and spectral information \cite{eyben2015geneva}. These knowledge-driven features are fed into decision-making algorithms, often supported by machine learning models, that classify emotions expressed in spoken utterances. With recent advances and breakthroughs in deep learning, especially Transformer frameworks \cite{vaswani2017attention}, a large number of works have focused on utilizing large-scale pre-trained speech models for SER leading to promising classification results on numerous existing SER benchmarks \cite{chen2021exploring, pepino21_interspeech, wagner2022dawn, feng2023trustser}.

For example, \cite{pepino21_interspeech} was one of the first SER studies to investigate fine-tuning with Wav2vec 2.0 embeddings. This paper proposed a framework that combines the output of each encoder layer from pre-trained Wav2vec models using trainable weights, which are learned jointly with the downstream model. This paper reported significantly better performance than previous works that rely solely on multilayer perceptrons (MLP), convolutional neural networks (CNNs), and recurrent neural networks (RNNs) \cite{zhao2019speech, feng2022enhancing}. In addition to experimenting on speech embeddings, \cite{chen2021exploring} proposed task-adaptive pretraining (TAPT) based on the Wav2vec 2.0 model, which further improves SER performance. Furthermore, \cite{wagner2022dawn} conducted a comprehensive analysis of fine-tuning Wav2vec 2.0 and HuBERT \cite{hsu2021hubert} for SER that also evaluates robustness, fairness, and efficiency.

\begin{table*}[t]
\caption{Comparison between our work and existing SER studies that investigate fine-tuning with pre-trained architectures.}
    \small
    \begin{tabular*}{\linewidth}{lcccccc}
        \toprule
        
        & \multirow{1}{*}{\shortstack{\textbf{Downstream FT}}} & 
        \multirow{1}{*}{\textbf{Adapter}} & 
        \multirow{1}{*}{\textbf{Embedding Prompt}} & 
        \multirow{1}{*}{\textbf{LoRa}} & 
        \multirow{1}{*}{\shortstack{\textbf{Pre-trained} \textbf{Architectures}}} &
        \multirow{1}{*}{\textbf{\#Datasets}} \\

        \midrule 
        \textbf{Chen et al.} \cite{chen2021exploring} & \cmark & \xmark & \xmark & \xmark & Wav2vec 2.0 Base & 2 \\

        \textbf{Pepino et al.} \cite{pepino21_interspeech} & \cmark & \xmark & \xmark & \xmark & Wav2vec 2.0 Base & 2 \\

        \multirow{1}{*}{\textbf{Wagner et al.}\cite{wagner2022dawn}} & \multirow{1}{*}{\cmark} & \multirow{1}{*}{\xmark} & \multirow{1}{*}{\xmark} & \multirow{1}{*}{\xmark} & Wav2vec 2.0 Families, HuBERT & 3 \\

        \textbf{Li et al.} \cite{li2023exploration} & \cmark & \xmark & \xmark & \xmark & Wav2vec 2.0 Base & 2 \\

        \midrule 
        \multirow{2}{*}{\textbf{Ours}} & 
        \multirow{2}{*}{\cmark} & \multirow{2}{*}{\cmark} & \multirow{2}{*}{\cmark} & 
        \multirow{2}{*}{\cmark} & \multirow{2}{*}{\shortstack{Wav2vec 2.0 Base, WavLM Base+\\ Whisper Tiny, Base, Small}} & 
        \multirow{2}{*}{\textbf{4}} \\
        
        & & & & & & \\ 
        \bottomrule
    \end{tabular*}
    \label{table:comparison_trust_ser}
    \vspace{-2mm}
\end{table*}

However, most existing works on the use of pre-trained speech models for SER rely heavily on fine-tuning all the model parameters (full fine-tuning) or training the downstream architecture (pre-trained model parameters are frozen), and approaches like parameter efficient fine-tuning (PEFT) have not been explored extensively for SER \cite{houlsby2019parameter}. PEFT is the prevalent methodology to adapt pre-trained language models (PLMs) to downstream tasks, providing strong performances on many popular NLP benchmarks without modifying the pre-trained architecture \cite{houlsby2019parameter}. Compared to conventional methods that fine-tune the entire pre-trained model, PEFT prevents storing separate copies of model parameters for individual downstream tasks, significantly reducing the required computational resources where modern pre-trained language models predominantly come with hundreds of millions of parameters or even hundreds of billions of parameters \cite{he2021towards}. The Speech UndeRstanding Evaluation (SURE) benchmark \cite{li2023evaluating} is one study closest to our work that explores PEFT for SER. However, the primary focus of SURE is general speech understanding, and their experiments cover only one pre-trained speech model and  limited SER datasets, providing limited knowledge about the efficacy of PEFT to SER.


Most PEFT methods involve updating a small number of extra parameters while leaving the existing pre-trained architecture unmodified. One such approach in this field is \texttt{adapter tuning} \cite{houlsby2019parameter}, which entails inserting small neural network modules into each transformer layer. These adapter modules usually comprise down-projection, up-projection, and the non-linear layer between projection layers. During adapter tuning, the only updated parameters are those in the inserted adapters, while the pre-trained model parameters stay unchanged. Another popular technique is \texttt{embedding prompt learning} \cite{jia2022visual}, which has shown promising results in various NLP tasks. This technique introduces prefix embeddings to the hidden layers that can be trained during fine-tuning. However, previous studies have shown that optimizing prefix embedding prompts can be challenging due to non-monotonic performance changes when varying trainable parameters. On the other hand, adding adapters can reduce inference efficiency. In more recent work, researchers have proposed \texttt{LoRa (Low-rank Adaptation)} \cite{hu2021lora}, which uses low-rank matrices to approximate model updates during the training stage, achieving both lower inference latency and ease of optimization.

In this work, we introduce PEFT-SER, which explores parameter-efficient transfer learning in SER, integrating \texttt{adapter tuning}, \texttt{embedding prompt learning}, and \texttt{LoRA}. We present a comprehensive analysis of four widely-used SER testbeds: IEMOCAP \cite{busso2008iemocap}, MSP-Improv \cite{busso2016msp}, MSP-Podcast \cite{lotfian2017building}, and CREMD-D \cite{cao2014crema}. Specifically, our contributions are summarized as follows:

\begin{itemize}[leftmargin=*]
    \item A novel exploration study for parameter-efficient transfer learning on pre-trained speech models for SER: \textbf{PEFT-SER}. 
    
    \item Comprehensive experiments cover \textbf{4 popular SER testbeds} using \textbf{5 representative pre-trained backbones}: Whisper Tiny, Base, Small \cite{radford2022robust}, Wav2vec 2.0 \cite{baevski2020wav2vec}, and WavLM \cite{chen2022wavlm}.

    \item Detailed evaluation of parameter-efficient fine-tuning methods using the \textbf{adapter}, \textbf{embedding prompt tuning}, and \textbf{LoRa}. Our results demonstrate that LoRa yields consistently better performance across all pre-trained architectures.
    
    \item An evaluation of the trustworthiness of using parameter-efficient transfer learning approaches, offering insights into balancing system performance, the number of parameters to add, and fairness.
    
\end{itemize}

\section{Background}

\subsection{Pre-trained Speech Architecture}

Self-supervised learning (SSL) is a widely-used machine learning paradigm that involves modeling and classifying input data without requiring explicit (human) labeling. SSL has gained considerable popularity in the field of NLP, where the models are commonly trained by reconstructing masked input tokens or predicting the next token in a given sentence. This approach has also shown great promise for speech representation learning, enabling training large-scale speech corpus without labels. Inspired by the learning objectives used in NLP, SSL methods on speech data frequently target learning generic speech representations by using generative~\cite{chen2022wavlm}, discriminative~\cite{baevski2020wav2vec,schneider2019wav2vec} and multi-task learning objectives~\cite{chen2022wavlm}. Relatedly, \cite{radford2022robust} proposed to use weakly-supervised learning approaches that train the transformer architecture on 680,000 hours of audio data, achieving state-of-the-art performance in automatic speech recognition tasks (ASR). Table~\ref{table:pretrained_models} summarizes the details of the pre-trained models used in this work.

\begin{table}[t]
\caption{Summary of pre-trained encoders used in this work.}

    \small
    \begin{tabular*}{\linewidth}{lcccc}
        \toprule
        
        \multirow{2}{*}{\shortstack{\textbf{Pre-trained}\\\textbf{Architecture}}} & 
        \multirow{2}{*}{\shortstack{\textbf{Input}\\\textbf{Data}}} & 
        \multirow{2}{*}{\shortstack{\textbf{\#Layers}}} &
        \multirow{2}{*}{\shortstack{\textbf{Hidden}\\\textbf{Size}}} & 
        \multirow{2}{*}{\shortstack{\textbf{\#Params}}}  \\ 

        & & & & \\ 
         
        \midrule
        \textbf{Whisper Tiny} & Mel Spec & 4 & 376 & 8.21M \\ 
        \textbf{Whisper Base} & Mel Spec & 8 & 512 & 20.59M \\ 
        \textbf{Whisper Small} & Mel Spec & 12 & 768 & 88.15M \\
        \textbf{W2V 2.0 Base} & Raw Wave & 12 & 768 & 95.04M \\ 
        \textbf{WavLM Base+} & Raw Wave & 12 & 768 & 94.70M \\ 

        \bottomrule
    \end{tabular*}
    \vspace{-3mm}
    \label{table:pretrained_models}
\end{table}

\noindent \textbf{Wav2vec 2.0} \cite{baevski2020wav2vec} is a transformer model \cite{vaswani2017attention} that utilizes a masked learning objective to predict true quantized latent speech representations from the remaining context. Wav2vec 2.0 takes raw speech signals as input.

\noindent \textbf{WavLM} \cite{chen2022wavlm} expands on the Wav2vec 2.0 pre-training objectives by incorporating masked speech denoising and frame prediction. This model achieves competitive performance on popular speech-based downstream tasks, such as speaker recognition, speaker diarization, and speech recognition.

\noindent \textbf{Whisper} \cite{radford2022robust} is trained with weakly supervised learning objectives, including VAD, language detection, ASR, etc. Fine-tuning Whisper for SER is an area of limited exploration.

\begin{figure*}[t]
    \begin{center}
        \includegraphics[width=0.85\linewidth]{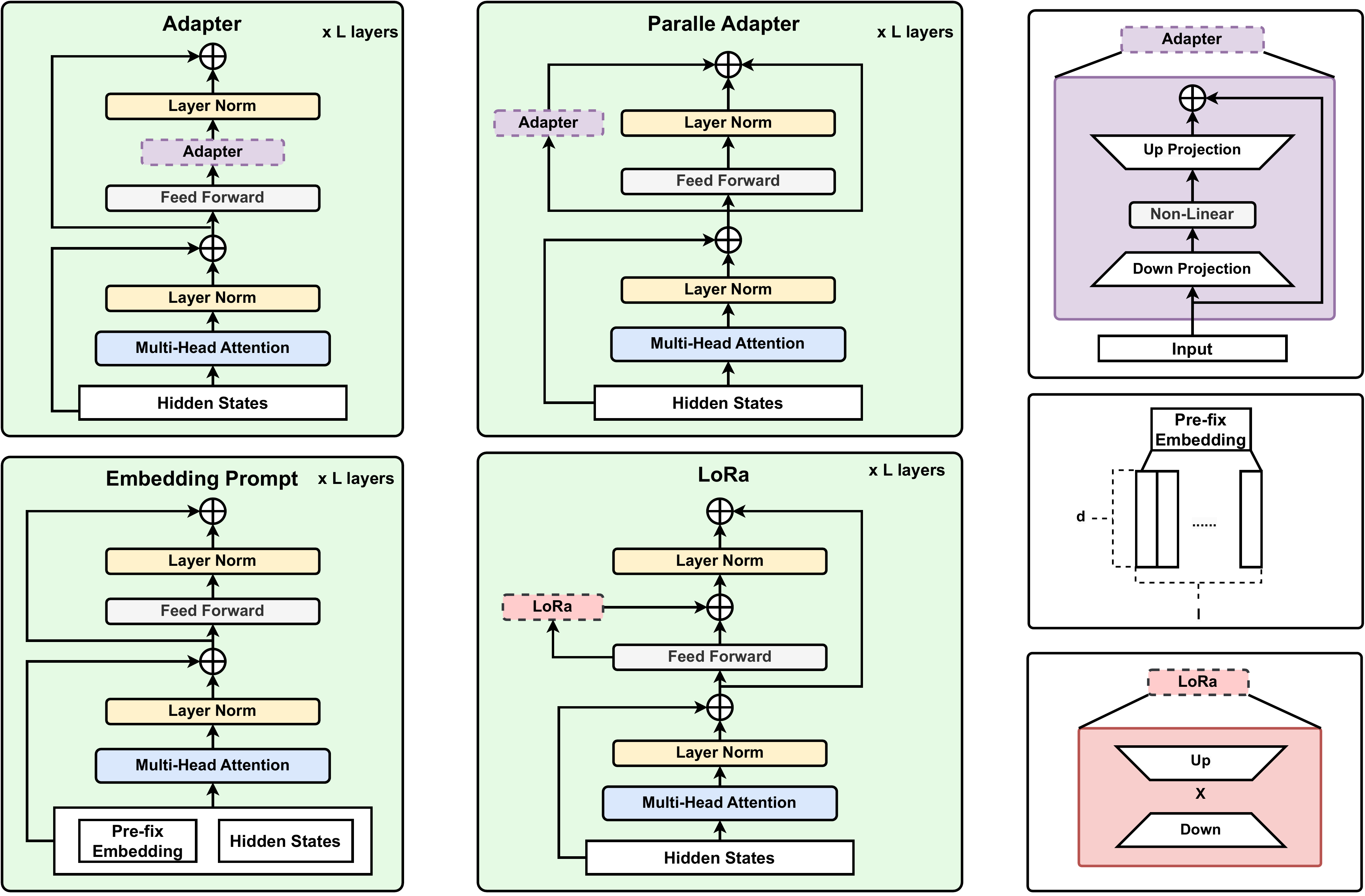}
    \end{center}
    \caption{System architecture of different parameter-efficient fine-tuning (PEFT) approaches used in this study.}
    \label{fig:peft}
    \vspace{-3mm}
\end{figure*}

\subsection{Fine-tuning Pre-trained Speech Models for SER}

As summarized in Table~\ref{table:comparison_trust_ser}, many SER frameworks that target fine-tuning existing pre-trained models have been developed in the past few years. For example, \cite{pepino21_interspeech} was one of the first SER studies to investigate fine-tuning with Wav2vec 2.0 embeddings. They have reported competitive results compared to previous literature that did not employ pre-trained models. 
Meanwhile, \cite{chen2021exploring} proposed task-adaptive pretraining (TAPT) based on the Wav2vec 2.0 model, which further improves SER performance. 
More recently, \cite{wagner2022dawn} conducted a comprehensive analysis of fine-tuning Wav2vec 2.0 and HuBERT \cite{hsu2021hubert} for SER. More recently, \cite{li2023exploration} studied the embedding importance to SER using Wav2Vec 2.0 models. Most above studies focus on fine-tuning the pre-trained architecture or pre-trained speech embeddings, none of these studies include the exploration of PEFT in their experiments. In contrast, this work presents a comprehensive analysis of using PEFT on pre-trained speech models for SER. Additionally, we explore the fine-tuning of WavLM and Whisper models, which have not been extensively studied in the field.

\subsection{Parameter-efficient Tuning Methods}

In this section, we provide details on the PEFT methods experimented on in this paper: adapter, embedding prompt tuning, and LoRa. We want to stress that pre-trained architecture is frozen in experiments. A summary of the PEFT methods is also provided in Figure \ref{fig:peft}.

\noindent \textbf{Adapter}
As mentioned in the previous section, the adapter approach introduces additional trainable layers within each transformer layer, including the down-project layer, the up-projection layer, and the non-linear layer in between \cite{houlsby2019parameter}. For instance, starting with an input vector $\mathbf{h}$ of embedding size $d$, the down-projection layer first outputs a lower-dimensional representation of size $e$. The resulting low-dimensional embeddings are then passed through a non-linear function $f(\cdot)$, such as ReLu, before being up-projected back to their original shape by the up-projection layer. Additionally, adapters frequently integrate a residual connection to the final output. Given the down-projection layer $\mathbf{W_{d}}$ and up-projection layer $\mathbf{W_{u}}$, the adapter includes the following computation:

\begin{equation}
    \mathbf{h} = \mathbf{h} + f(\mathbf{W_{d}} \mathbf{h}) \mathbf{W_{u}}
\end{equation}

Notably, there are many variants for integrating adapters. In this work, we explore the adapter in two forms, one connects the adapter to the output of the feed-forward layers (\textbf{Adapter}), while the other inserts the adapters as parallel components to feed-forward and layer-norm modules (Parallel Adapter). 

\noindent \textbf{Embedding Prompt Tuning:}
Drawing upon textual prompting techniques, embedding prompt tuning adds $l$ trainable embeddings to the input embedding space before each encoder layer \cite{jia2022visual}. The resulting prompt output from each preceding layer is removed, and a new set of prompts is concatenated before being fed into the subsequent layer. During downstream tasks,  only the parameters of the embedding prompts and classification layers are optimized for downstream tasks. The computation for each encoder layer in embedding prompt tuning, given an embedding prompt $\mathbf{e}$, is demonstrated below:

\begin{equation}
    \_, \mathbf{h} = \text{Enc} (\text{concat} (\mathbf{e}, \mathbf{h}))
\end{equation}

\noindent \textbf{LoRa:}
The LoRa approach proposes to fine-tune low-rank matrices to approximate the model updates \cite{hu2021lora}. For instance, considering a pre-trained linear layer $\mathbf{W}\in \mathcal{R}^{d \times k}$, where $d$ and $k$ are input and output dimensions, LoRa replaces model updates with a low-rank matrix decomposition:

\begin{equation}
    \mathbf{W} - \mathbf{\Delta W} = \mathbf{W} - \mathbf{W_{d}\mathbf{W_{u}}}
\end{equation}

where $\mathbf{W_{d}} \in \mathcal{R}^{d \times r}$ and $\mathbf{W_{u}} \in \mathcal{R}^{r \times k}$ where $r$ represents the low rank order. While this low-rank approximation can be applied to both attention and feed-forward layers, we explore LoRa, which applies to feed-forward layers.

\begin{figure}[t]
    \begin{center}
        \includegraphics[width=0.96\linewidth]{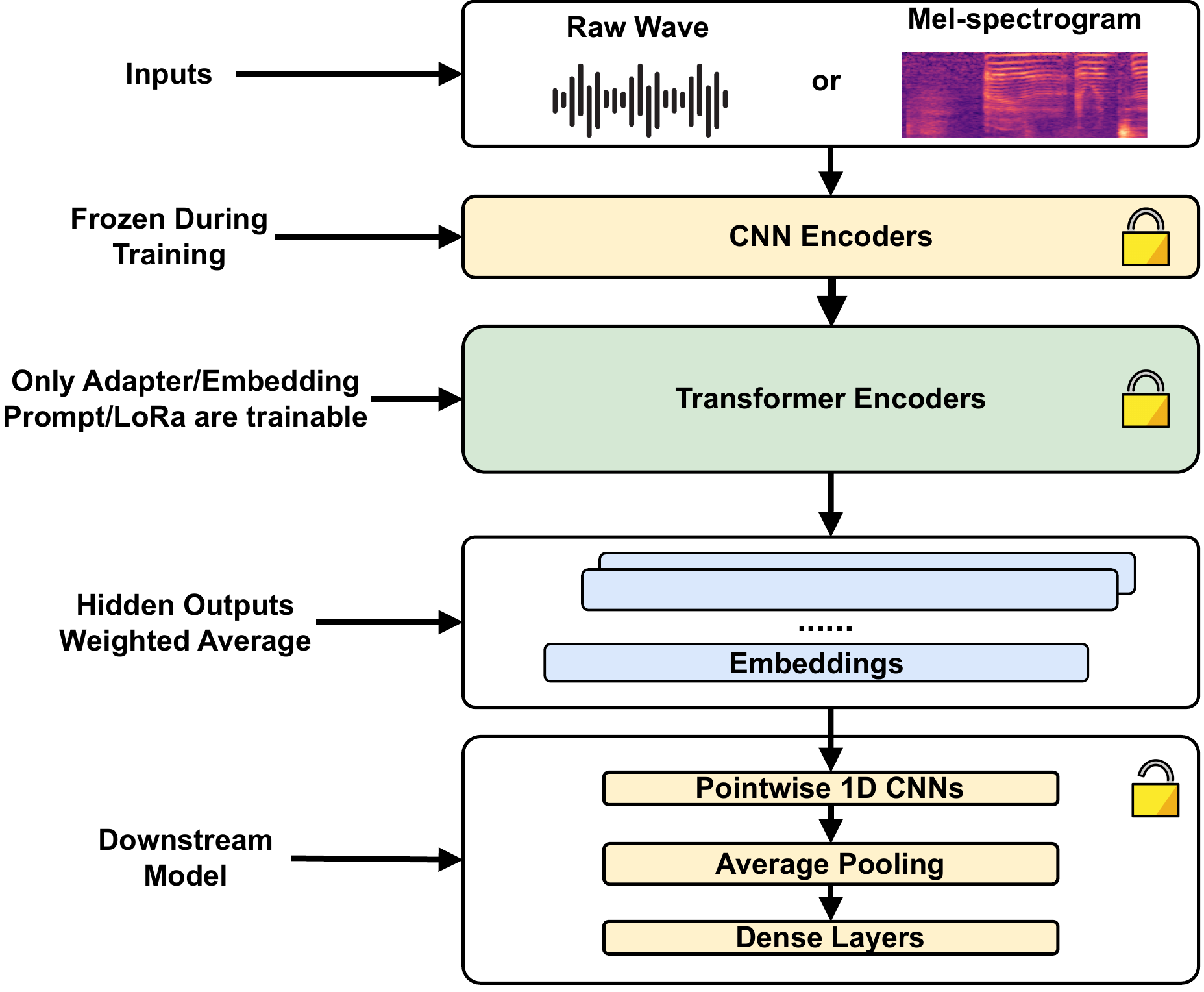}
    \end{center}
    \caption{Modeling framework used in this work. The pre-trained models shown in the diagram include Wav2vec 2.0 Base, WavLM Base+, and Whisper models.}  
    \label{fig:downstream_model}
    \vspace{-3mm}
\end{figure}

\section{Modeling Approach}

\noindent \textbf{Fine-tune Strategies}: In our experimentation, we independently apply the adapter, parallel adapter, embedding prompt, and LoRa to each pre-trained architecture. Moreover, we attempt to build baseline fine-tuning approaches to the above parameter-efficient fine-tuning methods. Inspired by the modeling results presented in \cite{pepino21_interspeech}, we identify that fine-tuning the pre-trained speech model performs poorly compared to freezing the backbone encoder in SER classification. Consequently, keeping the backbone encoder frozen provides simple but competitive results in SER applications. We want to highlight that we adopt the same downstream classification architecture in all fine-tuning approaches. In summary, we experiment with freezing the pre-trained model for fine-tuning, adapter tuning, embedding prompt tuning, and LoRa.

\vspace{1.5mm}
\noindent \textbf{Downstream Modeling}: We adopt the architecture proposed in \cite{pepino21_interspeech} for our downstream model, starting with weighted averaging to combine the hidden outputs of all encoder layers, where the weights are trainable parameters. Subsequently, we pass the weighted average output to three 1D point-wise convolutional layers with a filter size of 256 and a kernel size of 1, with ReLU activation functions in between. The outputs from the convolutional layers are then averaged over the timestamps which leads to a vector of size 256. Finally, this vector is fed into two fully connected layers for SER prediction. The details of the modeling architecture are provided in Fig~\ref{fig:downstream_model}\footnote{The figure uses images from https://openmoji.org/.}.

\section{Datasets}

In this section, we provide a brief overview of each included dataset and its corresponding recording condition. Table~\ref{table:datasets} shows emotion label statistics for the four datasets included in our PEFT experiments for SER. All are publicly available and widely used in the SER literature, ensuring accessibility and reproducibility. Due to the imbalanced label distribution within most SER datasets, we keep the four most frequently presented emotions from all experimental datasets, as recommended in \cite{feng2022enhancing, chen2021exploring, pepino21_interspeech}: neutral, happy, sad, and angry.

\vspace{1.5mm}
\noindent \textbf{IEMOCAP} database \cite{busso2008iemocap} comprises multi-modal recordings capturing motion, audio, and video of acted human interactions. The data are from ten subjects, evenly distributed between males and females, to express categorical emotions. The experimental data contains 5,531 utterances.

\vspace{1.5mm}
\noindent \textbf{CREMA-D} dataset consists of audio-visual clips that were recorded using 91 actors \cite{cao2014crema}. The participants in the study were directed to express six specific emotions while uttering a set of 12 sentences. The emotions targeted for expression were neutral, happy, angry, disgusted, fearful, and sad. The filtered dataset with four emotions contains 4,798 utterances.

\vspace{1.5mm}
\noindent \textbf{MSP-Improv}~\cite{busso2016msp} corpus is developed with the target of investigating naturalistic emotions that were elicited from improvised situations. The corpus is comprised of both audio and visual data captured during natural, target, improvised, and read speech conditions. The dataset was collected from 12 individuals, with an equal number of subjects from both male and female participants. To deeply understand the PEFT performance in different recording scenarios, we consider including utterances in all recording conditions (natural, target, improvised, and read). The number of utterances utilized in this particular study was 7,798.

\vspace{1.5mm}
\noindent \textbf{MSP-Podcast}~\cite{lotfian2017building} is collected from podcast recordings. We used the standard splits for training, validation, and testing (test set 1). This dataset has more than 500 speakers in the training split, 44 (22 female, 22 male) in the development, and 60 (30 female, 30 male) in the test split. More details on this dataset can be found in \cite{lotfian2017building}. We use the dataset release 1.8 (Oct. 26th, 2020) for the experiment.

\begin{table}[t]
\caption{Summary of dataset statistics used in this work.}
    \small
    \begin{tabular*}{\linewidth}{lccccc}
        \toprule
        
        \multirow{2}{*}{\shortstack{\textbf{Datasets}}} & 
        \multirow{2}{*}{\textbf{Neutral}} & 
        \multirow{2}{*}{\shortstack{\textbf{Happy}}} &
        \multirow{2}{*}{\shortstack{\textbf{Sad}}} & 
        \multirow{2}{*}{\shortstack{\textbf{Angry}}} & 
        \multirow{2}{*}{\shortstack{\textbf{Total}}}  \\ 

        & & & & & \\ 
         
        \midrule
        \textbf{IEMOCAP} & 1,708 & 1,636 & 1,084 & 1,103 & 5,531 \\ 
        \textbf{CREMA-D} & 1,972 & 1,219 & 588 & 1,019 & 4,798 \\ 
        \textbf{MSP-Improv} & 3,477 & 2,644 & 885 & 792 & 7,798 \\ 
        \textbf{MSP-Podcast} & 20,986 & 12,060 & 2,166 & 2,712 & 3,7924 \\
        \midrule
        \textbf{Total} & 28,143 & 17,559 & 4,723 & 5,716 & 56,051 \\

        \bottomrule
    \end{tabular*}
    \label{table:datasets}
    \vspace{-3mm}
\end{table}

\begin{figure*}[t]
    \begin{center}
        \includegraphics[width=\linewidth]{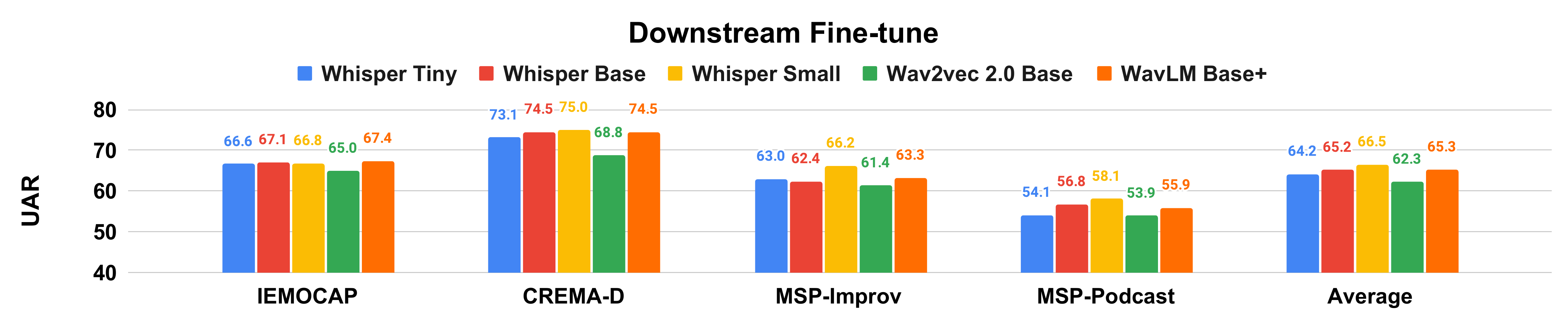}
    \end{center}
    \vspace{-3mm}
    \caption{Performance with fine-tuning downstream classification model (pre-trained model frozen during training) for SER.}
    \vspace{-3mm}
    \label{fig:downstream}
\end{figure*}

\section{Results}
\subsection{Experiment Details}

\noindent \textbf{Data Split}: We performed the speaker-independent evaluation to assess the quality of the trained models. We apply 5-fold and 6-fold evaluation on IEMOCAP and MSP-Improv datasets, where each session is regarded as a unique test fold. During each training fold, one session of data is used for validation while the rest are used for training. Moreover, we perform 5-fold cross-validation experiments on the CREMA-D dataset, where 20\% of speakers are used as testing data in each fold. On the other hand, we used the standard splits for training, validation, and testing from the MSP-Podcast dataset and repeated the training three times with different seeds. In summary, we perform 19 unique trainings on downstream finetuning and each PEFT approach to report the final results.

\vspace{1.5mm}
\noindent \textbf{SER Training}: We set the batch size as 32 and used the fixed seed in all fine-tuning experiments. Specifically, we set the learning rate as 0.0005 and the maximum training epoch as 30. We set the maximum audio duration as 6 seconds. The training input audio is augmented by noise addition and time masking. Specifically, we apply the Gaussian noise with SNR in the range of 10 and 30 dB and a time masking ratio between 10\% and 15\%. All experiments are implemented using PyTorch. The experiments are conducted on a high-performance computing server with A40 GPUs. We use the checkpoints of each pre-trained model from HuggingFace \cite{wolf-etal-2020-transformers}.

\vspace{1.5mm}
\noindent \textbf{Parameter Efficient Fine-tuning}: We set the bottleneck size in adapter tuning as 128 and the low-rank dimension in LoRa as 8. In embedding prompt learning, we choose an embedding prompt size of 5 in all experimental conditions. We summarize the total number of trainable parameters under the different fine-tuning conditions in Table~\ref{table:parameter_num}. It is worth noting that parameters of the downstream model should also be added to conditions of the adapter, embedding prompt, and LoRa.


\subsection{Downstream Model Performance}

In this section, we present the baseline results by fine-tuning downstream models while freezing the pre-trained encoders, as illustrated in Figure~\ref{fig:downstream}. From the results we can observe that large transformer models, such as Whisper Small and WavLM Base+, achieve the best overall performance across multiple SER datasets. Particularly, WavLM Base+ yields the best results on the IEMOCAP dataset, while Whisper Small outperforms the other models on the remaining datasets. In contrast, Wav2vec 2.0 does not provide competitive results compared to Whisper-Small and WavLM. Surprisingly, the plot indicates that the parameter-light Whisper-Tiny and Whisper-Base models produce better results than Wav2vec 2.0 despite having less than ~30M parameters in the pre-trained encoders. This finding highlights the possibilities of using smaller models in certain SER applications.

\begin{table}[t]
\caption{Summary of the number of trainable parameters under different settings.}

    \footnotesize
    \begin{tabular*}{\linewidth}{lcccc}
        \toprule
        
        \multirow{2}{*}{\shortstack{\textbf{Pre-trained}\\\textbf{Architecture}}} & 
        \multirow{2}{*}{\shortstack{\textbf{Downstream}\\\textbf{Model}}} &
        \multirow{2}{*}{\shortstack{\textbf{Adapter}}} &
        \multirow{2}{*}{\shortstack{\textbf{Embedding}\\\textbf{Prompt}}} & 
        \multirow{2}{*}{\shortstack{\textbf{LoRa}}} \\

        & & & & \\ 
         
        \midrule
        \textbf{Whisper Tiny} & 0.3 M & 0.40 M & 0.01 M & 0.06 M \\ 
        \textbf{Whisper Base} & 0.33 M & 0.79 M & 0.02 M & 0.12 M \\ 
        \textbf{Whisper Small} & 0.4 M & 2.37 M & 0.05 M & 0.37 M \\
        \textbf{W2V 2.0 Base} & 0.4 M & 2.37 M & 0.05 M & 0.37 M \\ 
        \textbf{WavLM Base+} & 0.4 M & 2.37 M & 0.05 M & 0.37 M \\ 

        \bottomrule
    \end{tabular*}
    \label{table:parameter_num}
    \vspace{-3mm}
\end{table}

\begin{table*}[t]
\caption{Performance comparisons between different PEFT methods for SER. The performance is denoted as $\mu\pm\sigma$.}
    \small
    \centering
    \begin{tabular*}{0.9\linewidth}{lccccc}
        \toprule
        
        & 
        \multicolumn{1}{c}{\textbf{Whisper Tiny}} &
        \multicolumn{1}{c}{\textbf{Whisper Base}} & 
        \multicolumn{1}{c}{\textbf{Whisper Small}} &
        \multicolumn{1}{c}{\textbf{Wav2vec 2.0 Base}} &
        \multicolumn{1}{c}{\textbf{WavLM Base+}} \\

        \midrule 
        \textbf{Downstream Tuning} & $64.21\pm7.92$  & $65.19\pm7.51$ & $\mathbf{66.53\pm6.91}$ & $62.27\pm6.32$ & $65.28\pm7.78$ \\
        
        \textbf{Adapter Tuning} & $57.80\pm7.26$  & $60.90\pm9.03$ & $64.36\pm7.96$ & $63.07\pm6.49$ & $\mathbf{67.09\pm9.35}$ \\

        \textbf{Paralle Adapter Tuning} & $55.18\pm6.36$  & $61.14\pm6.30$ & $61.62\pm7.91$ & $62.94\pm6.68$ & $\mathbf{65.87\pm9.30}$ \\

        \textbf{Embedding Prompt} & $57.42\pm6.93$  & $57.72\pm6.24$ & $58.40\pm6.56$ & $60.87\pm6.61$ & $\mathbf{66.66\pm8.79}$ \\

        \textbf{LoRa} & $63.42\pm7.02$  & $66.41\pm8.10$ & $66.54\pm9.50$ & $63.30\pm6.44$ & $\mathbf{67.28\pm8.79}$ \\

        \bottomrule
    \end{tabular*}
\vspace{-3mm}
\label{tab:peft_result}
\end{table*}

\subsection{Parameter-efficient Fine-tuning Performance}

We have further compared the overall performance (average UAR across four datasets) of PEFT methods with downstream classification baselines, as shown in Table~\ref{tab:peft_result}. Our findings indicate that adapter, parallel adapter, and embedding prompt perform worse than the direct downstream classification models when the pre-trained model is Whisper. Specifically, the parallel adapter approach exhibits the lowest performance among all parameter-efficient fine-tuning methods for the Whisper model. However, adapter, parallel adapter, and embedding prompt can lead to improved emotion recognition performance for Wav2Vec 2.0 Base and WavLM Base+ compared to training the downstream classification model alone.

Furthermore, LoRa consistently yields better SER performance when the pre-trained model is WavLM Base+ and Wav2Vec 2.0 Base. However, LoRa maintains relatively the same performance or even underperforms (e.g., Whisper Tiny) the downstream model training approach when applied to Whisper models. We conjecture this finding to be related to the positional embeddings in Whisper models, where future works should also consider finetuning the positional embeddings along with the PEFT. Our findings indicate that LoRa achieves the best fine-tuning performance (Average UAR across four datasets: $67.3\%$) on the WavLM Base+ model.

\begin{figure}[t]
    \begin{center}
        \includegraphics[width=\linewidth]{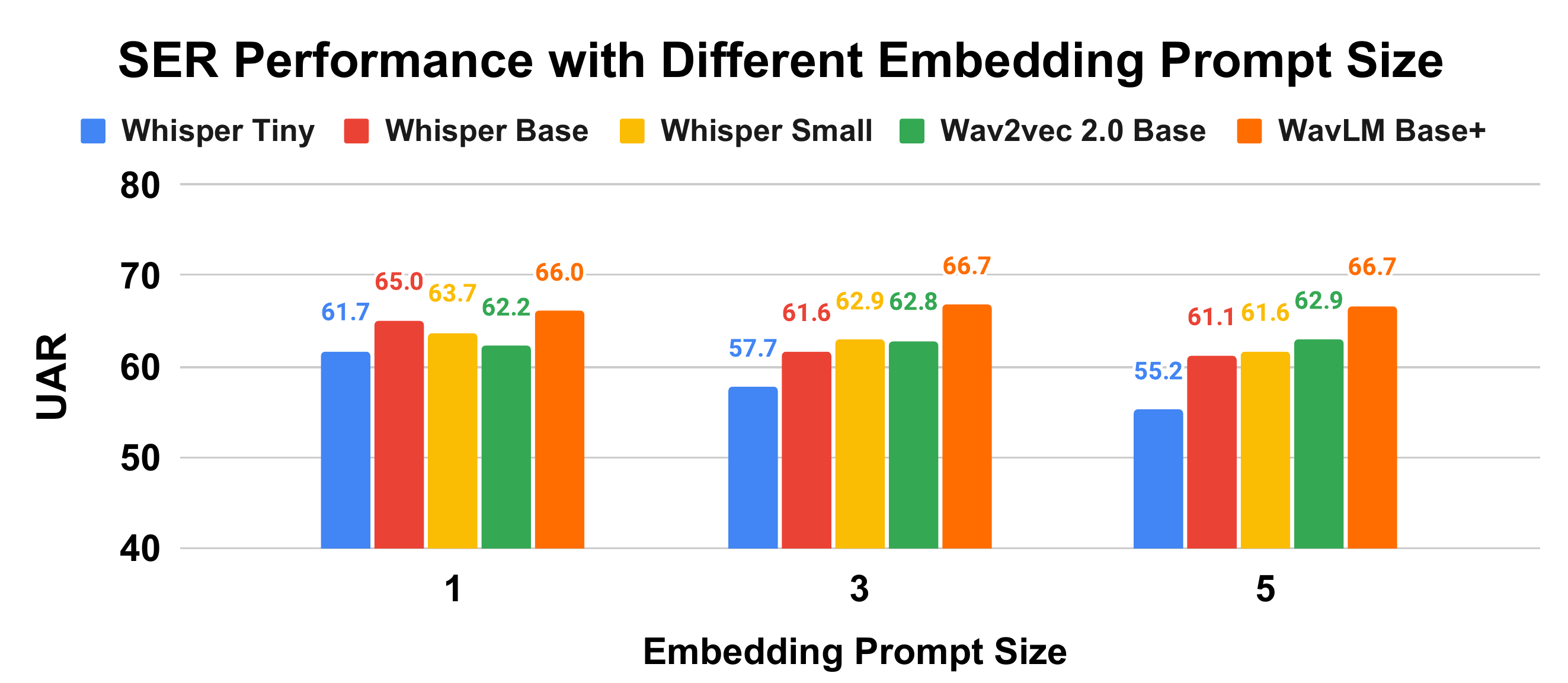}
    \end{center}
    \vspace{-3mm}
    \caption{SER performance varying embedding prompt sizes.}
    \label{fig:embedding_result}
    \vspace{-3mm}
\end{figure}

\section{Discussion}

In the previous section, we presented the results of the SER model performance using the parameter-efficient fine-tuning approach. Our findings show that the parameter-efficient fine-tuning approach consistently outperforms the downstream model-only approach on the WavLM Base+ model. However, it remains unclear whether and how the bottleneck size of the adapter, the number of embedding prompts, and the low-rank order in model update approximation impact the SER model performance. Additionally, it is unknown whether these hyper-parameters would have different influences on the performance of different pre-trained models. To answer these questions, we conduct several case studies in the following subsections.

\subsection{Impact of embedding prompt size on SER performance}

The findings in table~\ref{tab:peft_result} indicate that the embedding prompt approach results in significant subpar performance compared to fine-tuning the downstream classification model on all Whisper pre-trained models. However, whether changing the embedding prompt size directly impacts model performance remains unclear. To address this question, we conducted experiments varying the embedding prompt size $\in\{1, 3, 5\}$ during fine-tuning. The results of this experiment are presented in Figure~\ref{fig:embedding_result}. Our observations suggest that reducing the prompt embedding size enhances the performance of fine-tuning Whisper models for SER. Nevertheless, using an embedding prompt with a size of 1 still results in worse performance compared to the downstream model fine-tuning baseline. Additionally, we did not observe any direct impact of embedding size on the Wav2vec 2.0 base and WavLM base+ architecture, which aligns with prior research \cite{hu2021lora} that shows that fine-tuning embedding prompts is challenging, given that the model performance demonstrates non-monotonic patterns across different pre-trained architectures.

\begin{figure}[t]
    \begin{center}
        \includegraphics[width=\linewidth]{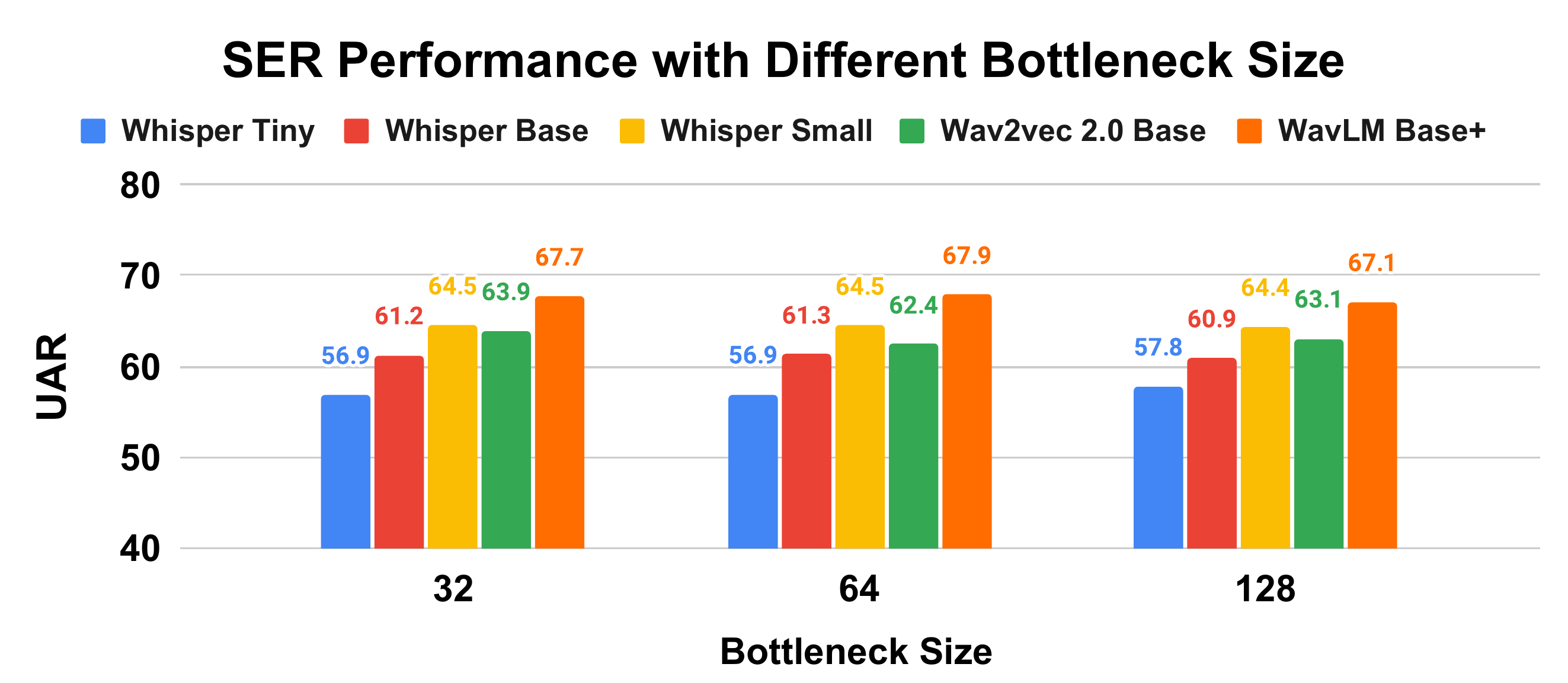}
    \end{center}
    \vspace{-3mm}
    \caption{SER performance with different bottleneck size.}
    \label{fig:adapter_result}
    \vspace{-3mm}
\end{figure}

\subsection{Impact of adapter bottleneck size on SER performances}
In this subsection, we investigate the association between the bottleneck size in adapter tuning and the SER performance in Figure~\ref{fig:adapter_result}. As parallel adapter tuning yields much worse performance than adapter tuning, we only discuss the results related to adapter tuning. Specifically, we experiment with bottleneck size $d \in \{32, 64, 128\}$.  Our results indicate that the bottleneck size has minimal impact on the SER performance under all conditions. This finding suggests that the adapter approach is robust to hyperparameter changes and can produce consistent results regardless of the pre-trained model architecture. However, the best-performed adapter tuning SER model still underperforms the LoRa approach.

\begin{table*}[t]
\caption{Fairness score on different parameter-efficient fine-tuning methods. Dem., Parity and Eq., of Odds represent demographic parity and equality of odds, respectively. The lower the score is, the better it is regarding fairness.}
    \scriptsize
    \centering
    \begin{tabular*}{0.93\linewidth}{lcccccccc}
        \toprule
        
        & 
        \multicolumn{2}{c}{\textbf{IEMOCAP}} &
        \multicolumn{2}{c}{\textbf{CREMA-D}} & 
        \multicolumn{2}{c}{\textbf{MSP-Improv}} &
        \multicolumn{2}{c}{\textbf{MSP-Podcast}} \\ 

        & 
        \multicolumn{1}{c}{\textbf{Dem., Parity}} &
        \multicolumn{1}{c}{\textbf{Eq., of Odds}} & 
        \multicolumn{1}{c}{\textbf{Dem., Parity}} &
        \multicolumn{1}{c}{\textbf{Eq., of Odds}} &
        \multicolumn{1}{c}{\textbf{Dem., Parity}} &
        \multicolumn{1}{c}{\textbf{Eq., of Odds}} &
        \multicolumn{1}{c}{\textbf{Dem., Parity}} &
        \multicolumn{1}{c}{\textbf{Eq., of Odds}} \\ 

        \midrule 
        \textbf{Downstream Tuning} & 12.25  & 13.83 & 7.97 & 11.45 & 13.15  & 23.49 & 14.96 & 20.51 \\
        
        \textbf{Adapter Tuning} & 11.79 & 11.50 & 7.89 & 11.08 & 13.34 & 21.66 & 15.98 & 18.52 \\

        \textbf{Embedding Prompt} & 12.01 & 12.79 & 9.07 & \textbf{9.97} & 13.32 & \textbf{21.11} & \textbf{13.14} & 17.49 \\

        \textbf{LoRa} & \textbf{10.98} & \textbf{11.14} & \textbf{7.48} & 11.43 & \textbf{12.92} & 22.43 & 14.71 & \textbf{15.96} \\

        \bottomrule
    \end{tabular*}
\vspace{-3mm}
\label{table:fairness}
\end{table*}

\subsection{Impact of low-rank order in LoRa on SER performance}

\begin{figure}[t]
    \begin{center}
        \includegraphics[width=\linewidth]{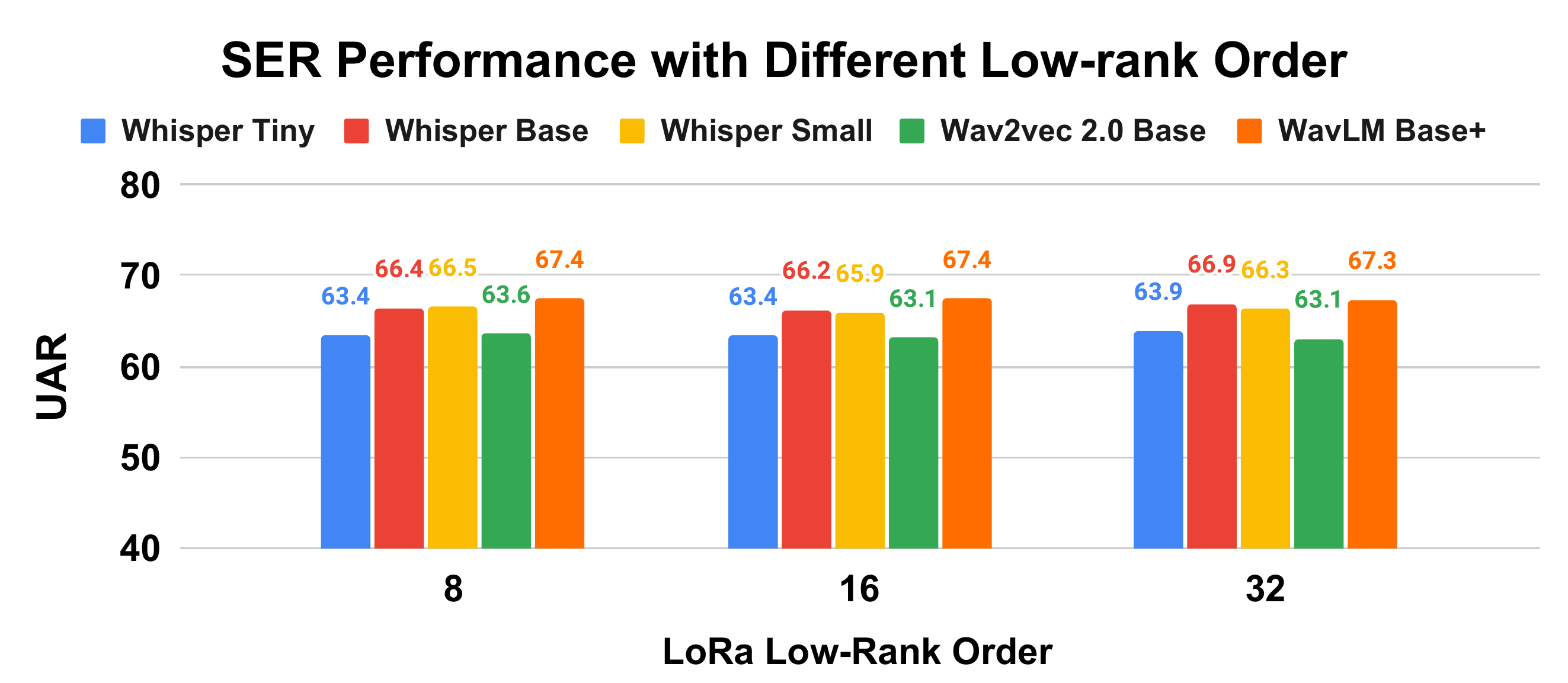}
    \end{center}
    \caption{SER performance with different low-rank orders.}
    \label{fig:lora_result}
    \vspace{-3mm}
\end{figure}

This subsection explores the relationship between the low-rank order in LoRa and its SER performance, as shown in Figure~\ref{fig:lora_result}. In particular, our experimentation involves varying the low-rank order within the set ${8, 16, 32}$. Our findings suggest that the SER performance is hardly impacted by changes in the low-rank order across all experimented conditions. This observation implies that, similar to adapter tuning, LoRa demands minimal hyperparameter tuning and can generate reliable outcomes irrespective of the pre-trained model architecture. Overall, LoRa achieves the best performance among all fine-tuning methods.

\subsection{LoRa yields the best SER performance, but is it fair?}

One major concern centering around deep learning is fairness. It has been widely known that pre-trained models can behave unfairly towards different demographics or individuals, often caused by the skewed data distribution presented in the pre-training dataset \cite{feng2023trust}. Such variable performance can lead to significant risks regarding discrimination and hinder the broader deployment of the service. Although LoRa outperforms other PEFT methods regarding system performance, there are still questions about whether it introduces biases in downstream predictions. To evaluate the fairness of different PEFT methods, we compute the equality of odds and demographic disparity across all datasets as suggested by \cite{gorrostieta2019gender}. 

\begin{figure}[t] {
    \centering
    
    \begin{tikzpicture}

        \node[draw=none,fill=none] at (0,0){\includegraphics[width=0.48\linewidth]{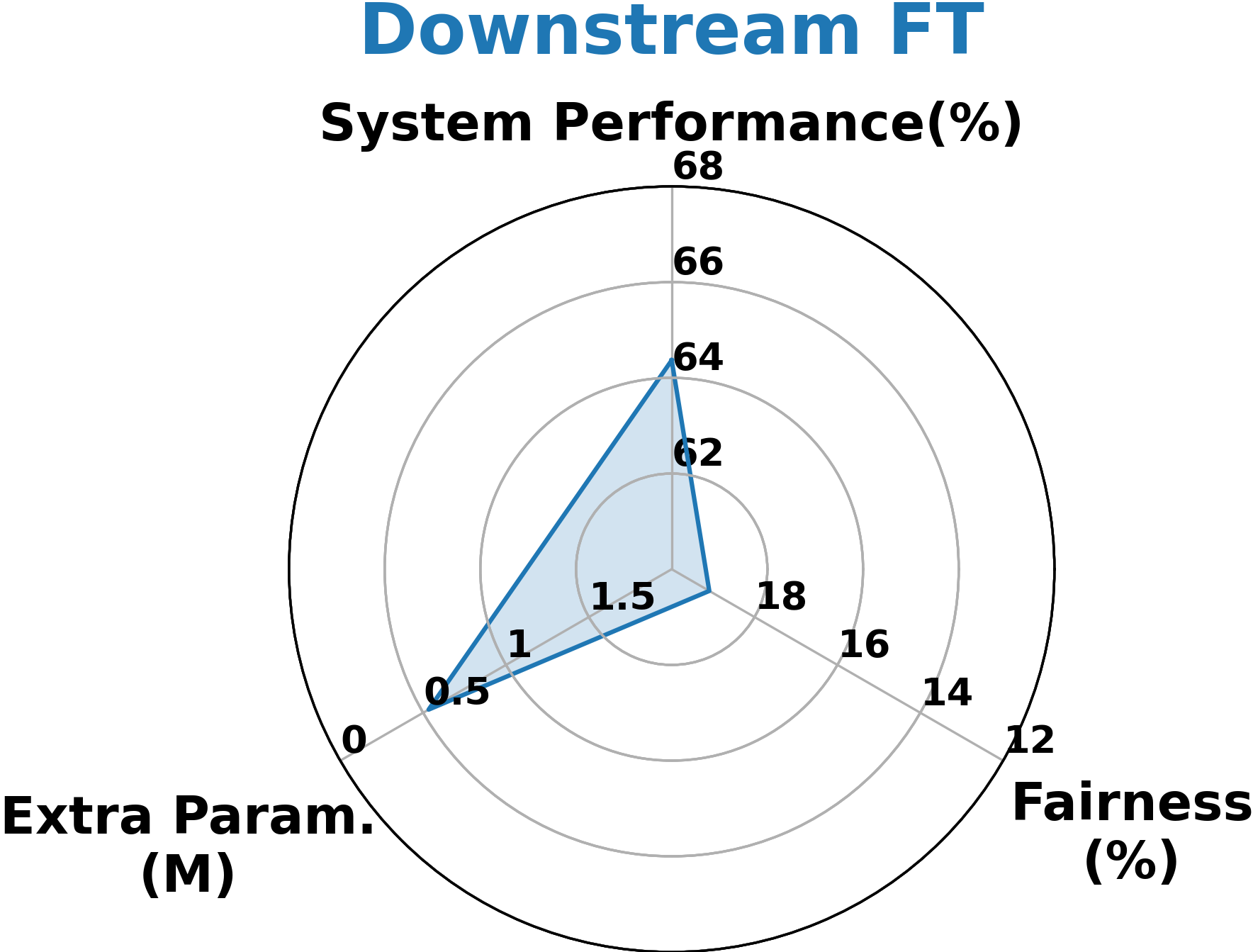}};

        \node[draw=none,fill=none] at (0.5\linewidth,0){\includegraphics[width=0.48\linewidth]{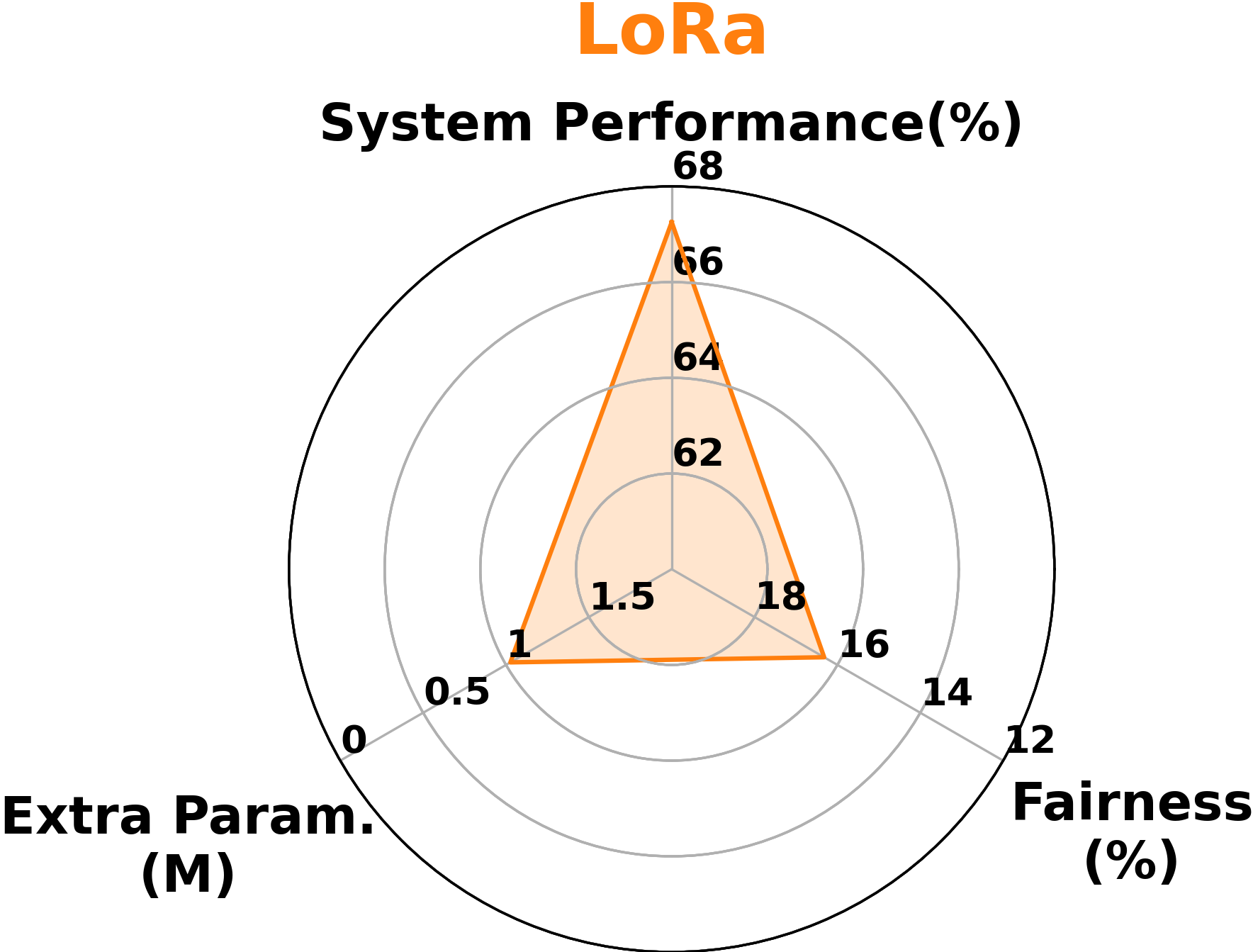}};
        
    \end{tikzpicture}
    \caption{Balances of system performance, extra parameters, and fairness between LoRa and downstream fine-tuning. The lower scores indicate better metrics for extra parameters and fairness.}
    \label{fig:trust_profile}
    \vspace{-3mm}
} \end{figure}

We report the results fine-tuned on WavLM Base+, as this model yields the best SER performance after applying PEFT. We use the low-rank order as 8 in LoRa, adapter size as 128, and embedding prompt size as 5, which provides the best results in individual PEFT. As shown in Table~\ref{table:fairness}, LoRa achieves the best fairness score in the majority of datasets, while the embedding prompt can also yield competitive fairness scores. It is worth noting that LoRa significantly outperforms downstream tuning across all datasets, highlighting the efficacy of PEFT in improving system performance and enhancing fairness for downstream tasks. The detailed comparisons between downstream fine-tuning and LoRa can be seen in Fig~\ref{fig:trust_profile}. Although our empirical evidence indicates that LoRa achieves better fairness than downstream tuning in SER, it is unclear whether this improvement is a benefit from the system performance increase. One possible future research is to investigate the fairness of PEFT approaches on more unbalanced training data regarding demographics.

\section{Conclusion}

In recent years, significant progress has been made in SER due to rapid advances in deep learning. While most existing works concentrate on the fine-tuning of pre-trained models for SER, there has been limited work investigating PEFT approaches for emotion recognition. To facilitate the research in this area, we conduct comprehensive experiments related to PEFT covering four popular SER datasets. Our findings demonstrate that LoRa achieves the best fine-tuning results on the WavLM Base+ model by introducing a minimal amount of learnable parameters. Furthermore, LoRa not only provides the best system performance but also yields the best fairness scores across most of the experimented datasets. Future research could explore the use of PEFT in multimodal emotion recognition applications.





\section*{Ethical Impact Statement}

With the rapid growth in deep learning, we have witnessed many promising pre-trained models that can capture general speech representation, substantially enhancing the performance for SER tasks. However, it is impractical to fine-tune the complete pre-trained model for different SER datasets collected, as it creates copies of the models that possibly include hundreds of millions, billions, or even trillion parameters, requiring enormous amounts of storage space. However, by adding several parameters, PEFT provides opportunities to redeploy the pre-trained speech models for SER. Increasingly, there are growing demands to deploy efficient and robust SER models in applications related to mobile computing and spatial computing (AR/VR). In these settings, it is critical to enable finetuning personalized SER models on edge devices. Given the limited computation resources on edge devices, it is more practical to utilize PEFT on pre-trained speech models to provide SER models that adapt to diverse environments. For example, PEFT creates the potential to apply Federated Learning for developing personalized SER \cite{zhang2023fedaudio, feng2022semi, feng2023fedmultimodal}.

 Meanwhile, the popularity of the foundation model \cite{bommasani2021opportunities}, typically built using a vast quantity of unlabeled data from the web, provide meaningful guides to perform future SER. However, fine-tuning foundation models is typically challenging due to the infrastructure limitations in academic settings. Alternatively, PEFT allows researchers to conduct more comprehensive research using publicly-available foundation models with an input of speech, offering the potential to facilitate understanding of using pre-trained speech models for SER. Lastly, PEFT allows ML practitioners to deploy SER models with unprecedented speed, while it is of particular importance to ensure the trustworthiness in applying PEFT for SER, including privacy breaches \cite{feng2022enhancing, feng2021privacy, jaiswal2020privacy}, unfair performance \cite{gorrostieta2019gender, chien2023achieving}, vulnerability to adversarial attacks \cite{goodfellow2014explaining}, and robustness of large pre-trained models \cite{leem2023adapting}. Finally, to facilitate research in SER using PEFT, we released model weights and packages that has been trained on the collection of dataset used in this work.

\bibliographystyle{IEEEtran}
\bibliography{ref}

\end{document}